\documentclass[prd,nofootinbib,onecolumn,preprint,11pt]{revtex4-2}

\usepackage{graphicx}
\usepackage{bm}
\usepackage{float}
\usepackage{subfigure}
\usepackage{amsmath}
\usepackage{amsfonts}
\usepackage{color}
\usepackage{slashed}
\usepackage{fancyhdr}
\usepackage[colorlinks=true,linkcolor=black,citecolor=black,filecolor=black,urlcolor=black,unicode]{hyperref}
\allowdisplaybreaks

\newcommand{\GN}{G_{\rm N}}
\newcommand{\gb}{\bar{g}}

\newcommand{\dd}{\text{d}}

\begin{document}

\title{Divergent Energy-Momentum Fluxes In Nonlocal Gravity Models}

\author{Yi-Zen Chu$^{1,2}$ and Afidah Zuroida$^1$}

\affiliation{
	$\,^1$Department of Physics, National Central University, Chungli 32001, Taiwan \\
	$\,^2$Center for High Energy and High Field Physics (CHiP), National Central University, Chungli 32001, Taiwan}

\begin{abstract}
\noindent We analyze the second order perturbations of the Deser-Woodard II (DWII), Vardanyan-Akrami-Amendola-Silvestri (VAAS) and Amendola-Burzilla-Nersisyan (ABN) nonlocal gravity models in an attempt to extract their associated gravitational wave energy-momentum fluxes. In Minkowski spacetime, the gravitational spatial momentum density is supposed to scale at most as $1/r^{2}$, in the $r \rightarrow \infty$ limit, where $r$ is the observer-source spatial distance. The DWII model has a divergent flux because its momentum density goes as $1/r$; though this can be avoided when we set to zero the first derivative of its distortion function at the origin. Meanwhile, the ABN model also suffers from a divergent flux because its momentum density goes as $r^{2}$. The momentum density from the VAAS model was computed on a cosmological background expressed in a Fermi-Normal-Coordinate system, and was found to scale as $r$. For generic parameters, therefore, none of these three Dark Energy models appear to yield well-defined gravitational wave energies, as a result of their nonlocal gravitational self-interactions.
\end{abstract}

\maketitle
\newpage
\tableofcontents

\section{Motivation and Introduction}
\label{Section_Introduction}

General Relativity (GR) has been well tested as a theory of gravity in numerous experiments and observations. Its physical predictions, such as the precession of Mercury's perihelion, gravitational lensing, and the existence of gravitational waves, have all been confirmed \cite{GRtest}. Meanwhile, cosmological observation data from supernovae has shown that our current universe's expansion is accelerating \cite{supernovae1, supernovae2}. A potentially unknown form of energy called dark energy can give rise to the late-time acceleration of our universe on cosmological scales. The simplest candidate for dark energy is the cosmological constant, symbolized as $\Lambda$ \cite{planck}. However the energy density of the cosmological constant estimated from quantum field theory appears to contradict the measurement from cosmology \cite{CCproblem}.

There have been various attempts to modify GR to explain the current phase of cosmic expansion without dark energy. In this work, we study nonlocal modifications of general relativity motivated by quantum loop effects. The nonlocal $1/\Box$, i.e., the inverse of the d'Alembertian operator, acting on the Ricci scalar $R$ is introduced to produce late-time cosmic accelerated expansion without dark energy. We focus on these three models of the nonlocal gravity, with actions given by
\begin{eqnarray}
	S_\text{DW II} &=& -\frac{1}{16 \pi \GN} \int \dd^{4} x \sqrt{-g} \left(1 + f[Y] \right) R, \label{eq:DWII} \\
	S_\text{ABN} &=&  -\frac{1}{16 \pi \GN} \int \dd^{4} x \sqrt{-g} \left(1 + \frac{M^{4}}{6} \frac{1}{\Box^{2}} \right) R, \label{eq:ABN} \\
	S_\text{VAAS} &=&  -\frac{1}{16 \pi \GN} \int \dd^{4} x \sqrt{-g} \left(1 + m^{2} \frac{1}{\Box} \right) R \label{eq:VAAS} ;
\end{eqnarray}
where $f[Y]$ is the distortion function of the nonlocal operator 
\begin{align}
	Y \equiv \frac{1}{\Box} \left( g^{\mu\nu} \partial_{\mu} \left( \frac{R}{\Box} \right) \partial_{\nu} \left( \frac{R}{\Box} \right) \right) .
\end{align}
(In the first model of Deser-Woodard \cite{DW I}, the form of the action is the same as that in eq. \eqref{eq:DWII}, but $f[Y]$ is replaced with $f[R/\Box]$.) The names of these models are acronyms formed from their respective authors in \cite{DW II, VAAS,  ABN}. For the nonlocal terms in equations \eqref{eq:ABN} and \eqref{eq:VAAS}, a new mass parameter has been introduced -- $M^{4}$ in the ABN and $m^{2}$ in the VAAS model -- that is of the same order as the cosmological constant $\Lambda$; namely, $\sqrt{\Lambda} \sim m \sim M$. The equations of motion corresponding to DWII, VAAS, and ABN all take the form
	\begin{align}
		\label{EoM}
		G_{\mu\nu} + \Delta G_{\mu\nu} = 8\pi\GN T_{\mu\nu} ,
	\end{align}
where $G_{\mu\nu}$ in the Einstein tensor; $\Delta G_{\mu\nu}$ are the nonlocal modifications to GR specific to each model which we will display in the later sections; and $T_{\mu\nu}$ is the stress tensor of matter.

Our primary goal is to study the energy-momentum flux of gravitational waves governed by the models in equations. \eqref{eq:DWII}, \eqref{eq:ABN}, and \eqref{eq:VAAS}. This extends the results of \cite{GWFluxDWI}, which only analyzed DWI \cite{DW I}. Now, the rate of energy per time through a sphere of radius $r$ emitted by an isolated source is given by 
\begin{eqnarray}
	\label{GWFlux}
	\left\langle \frac{\text{d}E}{\text{d}t} \right\rangle &=& 
	\lim_{r\to\infty} \int_{\mathbb{S}^{2}} \text{d}^{2} \Omega_{\widehat{r}} r^{2} \widehat{r}^{i} \langle t^{0i} \rangle ,
\end{eqnarray}
where $r$ denotes the distance from the source to the observer, $\widehat{r}$ is the unit radial vector, $\int_{\mathbb{S}^{2}} \text{d}^{2} \Omega_{\widehat{r}}$ is the associated integral over the solid angle, and $t^{0i}$ refers to the spatial momentum components of $t^{\mu\nu}$, the pseudo stress energy tensor of the gravitational waves. Thus, in (3+1)D Minkowski the gravitational energy-momentum flux $t^{0i}$ is supposed to scale exactly as $1/r^{2}$ in the far zone $r \to \infty$, so as to contribute a finite flux of energy to infinity. Moreover, in GR, this total power loss can be re-written in term of the source's quadrupole moment as 
\begin{eqnarray}
	\label{GRQuadrupoleRadiation}
	\left\langle \frac{\text{d}E}{\text{d}t} \right\rangle &=& \frac{\GN}{5} \left \langle \dddot{I}^{ij}_{tt}[t-r]  \dddot{I}^{ij}_{tt}[t-r]  \right \rangle  , 
\end{eqnarray}
where we have defined\footnote{Some words on index notation: Greek letters run from $0$ to $3$, where the zeroth component refers to time and the rest space. Latin/English alphabets run from the spatial dimensions $1$ through $3$. Einstein convention is in force unless otherwise stated.}
\begin{eqnarray}
	I_{tt}^{ij} &=& \frac{1}{2} \left( P^{i\{a} P^{b\}j} - P^{ij} P^{ab} \right) I^{ab}, \\
	P^{ij} &\equiv& \delta^{ij} - \widehat{r}^i \widehat{r}^j \\
	I^{ab}[t] &\equiv& \int_{\mathbb{R}^{3}} \text{d}^{3} \vec{x} (x-x_\text{CM})^{a} (x-x_\text{CM})^{b} T_{00}[t, \vec{x}] ,
\end{eqnarray}
for some spatial coordinate origin $\vec{x}_\text{CM}$. Eq. \eqref{GRQuadrupoleRadiation} is known as the quadrupole radiation formula. The discovery of the Hulse and Taylor binary pulsar in 1974 \cite{HulseTaylor} has led to the verification of this quadrupole radiation formula at the sub-percent level.

For the nonlocal models under study, we shall begin with an empty spacetime $\gb_{\mu\nu}$ with $T_{\mu\nu}=0$, which -- according to eq. \eqref{EoM} -- obeys
\begin{align}
	G_{\mu\nu}[\gb] + \Delta G_{\mu\nu}[\gb] = 0 .
\end{align}
We then proceed to insert an isolated source which perturbs the gravitational field $g_{\mu\nu} = \gb_{\mu\nu} + h_{\mu\nu}$. By expanding the equations-of-motion in eq. \eqref{EoM} up to second order in the metric perturbation $h_{\mu\nu}$, we obtain
	\begin{align}
		\delta_1 G_{\mu\nu} + \delta_1 \Delta G_{\mu\nu} = 8\pi\GN \left(\bar{T}_{\mu\nu} \left(1 + \mathcal{O}[h]\right) + \sum_{n=2}^{+\infty} \frac{\delta_n G_{\mu\nu} + \delta_n \Delta G_{\mu\nu}}{-8\pi\GN} \right) ,
	\end{align}
	with $\bar{T}_{\mu\nu}$ denoting the piece of the stress tensor independent of $h_{\mu\nu}$ and $\delta_n(\dots)$ denotes the piece of the tensor in $\dots$ that contains exactly $n$ powers of the metric perturbation. Moreover, as long as $h_{\mu\nu}$ remains small, we expect the second order terms to dominate over the higher ones $\delta_{n \geq 3} G_{\mu\nu} + \delta_{n \geq 3} \Delta G_{\mu\nu}$. One may also readily check that the divergence of the left hand side is identically zero; which implies, in the far zone where $T_{\mu\nu}=0$, the divergence of these second order terms must vanish at leading order in the $1/r$ expansion. These lead us to the usual interpretation that they, in fact, constitute the pseudo stress energy of the gravitational waves in these models:
	\begin{align}
		\label{eq:energy} 
		t_{\mu\nu} \equiv -\frac{\delta_{2} G_{\mu\nu} + \delta_{2} \Delta G_{\mu\nu}}{8\pi\GN} .
\end{align}
In \cite{GWFluxDWI}, the nonlocal interactions of DWI were shown to produce a $t_{\mu\nu}$ that went as $1/r$, which yields a gravitational wave (GW) flux in eq. \eqref{GWFlux} that diverges at infinity. By setting $f'[0] = f''[0] = 0$, however, the divergent flux can be avoided. We shall see that the DWII, VAAS, and ABN models also produce gravitational energy-momentum $t^{0i}$ that fall off slower than $1/r^2$.

This paper is organized as follows. In Section \eqref{Section_DWII}, we investigate DWII in a Minkowski background. We first derive its linearized solutions; then expand the nonlocal terms up to the quadratic order in metric perturbations, from which its gravitational wave energy-momentum is examined. Section \eqref{Section_ABN} shows that the ABN model can be analyzed in a similar fashion by expanding about flat spacetime; though its linearized solution already involves nonlocal interactions and can only be solved perturbatively up to first order in $M^4$. Finally, in section \eqref{Section_VAAS}, we explain why Minkowski is not an empty spacetime solution to the VAAS model. Instead, we shall expand about an empty cosmological background. Because its exact solution is difficult to obtain, however, we will solve it in the Fermi Normal Coordinate system, so that the metric itself is now the flat one plus corrections proportional to some power of the displacement $m \Delta \vec{X}$. Hence, we will have two expansions: one in powers of $h_{\mu\nu}$, and another in powers of $m^2$. We then perturb the nonlocal terms and the Einstein tensor up to the quadratic order in metric perturbations.

\section{Deser-Woodard II}
\label{Section_DWII}

After varying the action in eq. \eqref{eq:DWII} with respect to the metric, we obtain the equations of motion in eq. \eqref{EoM} with
\begin{align}
\Delta G_{\mu\nu} 
&= (G_{\mu\nu} + g_{\mu\nu} \Box -  \nabla_{\mu} \nabla_{\nu})(U + f[Y]) \nonumber\\
&+ \frac{1}{2}\partial_{\{ \mu} X \partial_{\nu\}} U + \frac{1}{2}\partial_{\{ \mu} Y \partial_{\nu \} } V 
+ V \partial_{\mu} X \partial_{\nu} X -  \frac{1}{2} g_{\mu\nu} g^{\rho\sigma} (\partial_{\rho} X \partial_{\sigma} U + \partial_{\rho} Y \partial_{\sigma} V + V \partial_{\rho} X \partial_{\sigma} X) ; \label{eq:eomDWII}
\end{align}
where the symmetrization symbol denotes, for e.g., $T_{\{\alpha\beta\}} \equiv T_{\alpha\beta} + T_{\beta\alpha}$. In this second model of Deser and Woodard, its nonlocal character is encoded within the distortion function $f[Y]$. To localize the theory, we have introduced the auxiliary fields $\{X,Y,V,U\}$, which obey
\begin{eqnarray}
	\label{eomDWII_X}
	\Box X &=& R, \\
	\label{eomDWII_Y}
	\Box Y &=& g^{\mu\nu} \partial_{\mu} X \partial_{\nu} X, \\
	\label{eomDWII_V}
	\Box V &=& R f'[Y], \\
	\label{eomDWII_U}
	\Box U &=& - 2 \cdot \nabla_{\mu} (V \nabla^{\mu} X). 
\end{eqnarray}
We use the metric signature $(+,-,-,-)$ and consider the perturbed flat spacetime metric
\begin{eqnarray}
	g_{\mu\nu} = \eta_{\mu\nu} + h_{\mu\nu} .
\end{eqnarray}
Throughout our analysis, we employ the trace-reverse perturbation defined via
\begin{align}
	h_{\mu\nu} &\equiv \bar{h}_{\mu\nu} - \frac{1}{2} \eta_{\mu\nu} \bar{h} , \\
	h &\equiv \eta^{\alpha\beta} h_{\alpha\beta} = -\eta^{\alpha\beta} \bar{h}_{\alpha\beta} \equiv -\bar{h} .
\end{align}
{\bf Auxiliary fields} \qquad Similar to the first model of Deser and Woodard in \cite{DW I}, Minkowski spacetime is an exact vacuum solution in this second model. Specifically, we may choose integration constants such that all the auxiliary fields in equations.\eqref{eomDWII_X}--\eqref{eomDWII_U} vanish in flat spacetime, where the Ricci scalar $R$ is zero. That $g_{\mu\nu}=\eta_{\mu\nu}$ is a solution to eq. \eqref{eq:eomDWII} for $T_{\mu\nu}=0$ may then be readily verified. We shall then expand the equations-of-motion up to second order in the (trace-reversed) metric perturbation $\bar{h}_{\mu\nu}$. Denoting $G[x,x']$ as the retarded Green's function $1/\Box$, the key ingredients involved are
\begin{eqnarray} 
	X 
	&=& \int \dd^4 x' \sqrt{|g[x']|} G[x,x'] R[x'] \\
	&\approx& \int_{x'} {}_{(1)}\overline{G}^{+} \delta_{1} R + \frac{1}{2} \int_{x'} h \ {}_{(1)}\overline{G}^{+}  \delta_{1} R + \int_{x'} \delta_{1} {}_{(1)}\overline{G}^{+} \delta_{1} R + \int_{x'} {}_{(1)}\overline{G}^{+} \delta_{2} R , \\ 
	Y   
	&=& \int \dd^4 x' \sqrt{|g[x']|} G[x,x'] \nabla_{\mu'} X \nabla^{\mu'} X \\
	&\approx& \int_{x'} {}_{(1)}\overline{G}^{+} \left( \partial^{\mu'} \int_{x''} {}_{(1)}\overline{G}^{+} \delta_{1} R \ \ \partial_{\mu'} \int_{x'''} {}_{(1)}\overline{G}^{+} \delta_{1} R \right) , \\
	f[Y] 
	&\approx& \bar{f} + \bar{f}' \int_{x'} {}_{(1)}\overline{G}^{+} \left( \partial^{\mu'} \int_{x''} {}_{(1)}\overline{G}^{+} \delta_{1} R \ \ \partial_{\mu'} \int_{x'''} {}_{(1)}\overline{G}^{+} \delta_{1} R \right) , \\ 
	f'[Y] 
	&\approx& \bar{f}' + \bar{f}'' \int_{x'} {}_{(1)}\overline{G}^{+} \left( \partial^{\mu'} \int_{x''} {}_{(1)}\overline{G}^{+} \delta_{1} R \ \ \partial_{\mu'} \int_{x'''} {}_{(1)}\overline{G}^{+} \delta_{1} R \right) , \\
	V 
	&=& \int \dd^4 x' \sqrt{|g[x']|} G[x,x'] R[x'] f[Y] \\
	&\approx& \bar{f}' \int_{x'} {}_{(1)}\overline{G}^{+} \delta_{1} R + \frac{\bar{f}'}{2} \int_{x'} h \ {}_{(1)}\overline{G}^{+}  \delta_{1} R + \bar{f}' \int_{x'} \delta_{1} {}_{(1)}\overline{G}^{+} \delta_{1} R + \bar{f}' \int_{x'} {}_{(1)}\overline{G}^{+} \delta_{2} R , \\ 
	U 
	&=& -2 \int \dd^4 x' \sqrt{|g[x']|} G[x,x'] \nabla_{\mu'} (V \nabla^{\mu'} X) \\
	&\approx& -2 \bar{f}' \int_{x'} {}_{(1)}\overline{G}^{+} \partial_{\mu'}  \left( \int_{x''} {}_{(1)}\overline{G}^{+} \delta_{1} R \partial^{\mu'} \int_{x'''} {}_{(1)}\overline{G}^{+} \delta_{1} R \right) ;
\end{eqnarray}
with $\bar{f} \equiv f[0]$ and $\bar{f}' \equiv f'[0]$. Here and below, we employ the shorthand $\int_{x'} \equiv \int_{\mathbb{R}^{3,1}} \dd^4 x'$ and $\int_{\vec{x}'} \equiv \int_{\mathbb{R}^{3}} \dd^3 \vec{x}'$. Note that the zeroth order $1/\Box$ is simply $1/\partial^2 = \,_{(1)}\overline{G}^+$, the Minkowski spacetime retarded Green function 
\begin{eqnarray}
{}_{(1)}\overline{G}^{+} [x-x'] = \frac{\delta [t - t' - |\vec{x} - \vec{x}'|] }{4 \pi |\vec{x} - \vec{x}'| } ; \label{eq:green} 
\end{eqnarray}
The first correction to the Green's function is \cite{greenfunction}
\begin{align}
\delta_1 {}_{(1)}\overline{G}^{+} [x,x']
= - \partial_\mu \partial_{\nu'} \int_{x''} {}_{(1)}\overline{G}^{+} [x-x''] \bar{h}^{\mu''\nu''} {}_{(1)}\overline{G}^{+} [x''-x'] . 
\end{align}
Moreover, the primes on the indices indicate the quantity is evaluated at $x'$, $x''$, etc.; for instance, $\partial_{\mu'} \equiv \partial/\partial x'^\mu$ and $\bar{h}^{\mu''\nu''} \equiv \bar{h}^{\mu\nu}[x'']$. Here and throughout the paper, we have also exploited, for arbitrary expansion $g_{\mu\nu} = \gb_{\mu\nu} + h_{\mu\nu}$ off the `background' $\gb_{\mu\nu}$, the expansions
\begin{align}
	g^{\mu\nu} &= \gb^{\mu\nu} - h^{\mu\nu} + h^{\mu}_{\phantom{\mu}\sigma} h^{\sigma\nu} + \dots \\
	\sqrt{|g|} &= \sqrt{|\gb|} \left( 1 + \frac{1}{2} h + \frac{1}{8} h^2 - \frac{1}{4} h_{\alpha\beta} h^{\alpha\beta} + \dots \right) ;
\end{align}
where $\gb^{\mu\nu}$ is the inverse of $\gb_{\mu\nu}$; $h \equiv \gb^{\mu\nu} h_{\mu\nu}$; and all indices on $h_{\mu\nu}$ are moved with $\gb_{\mu\nu}$.

\textbf{Linearized Solutions} \qquad To solve the GW energy-momentum in eq. \eqref{eq:energy} we first linearize eq. \eqref{eq:eomDWII}:
\begin{eqnarray}
	\label{DWII_Linearized}
	\delta_{1}G_{\mu\nu} (1 + \Bar{f}) = 8 \pi \GN \tau_{\mu\nu} ,
\end{eqnarray}
where $\tau_{\mu\nu}$ is, in principle, the stress tensor developed up to order $h$. Throughout the rest of this paper, we shall employ the de Donder gauge
\begin{eqnarray}
	\partial^{\mu} \bar{h}_{\mu\nu} = 0.
\end{eqnarray}
The first order of the Einstein tensor becomes 
\begin{eqnarray}
	\label{Einstein_1stOrder_deDonder}
	\delta_{1}G_{\mu\nu} &=& -\frac{1}{2} \partial^{2} \bar{h}_{\mu\nu}.
\end{eqnarray}
After employing eq. \eqref{Einstein_1stOrder_deDonder} in eq. \eqref{DWII_Linearized} we obtain
\begin{eqnarray}
	\bar{h}_{\mu\nu} &=& - \frac{16 \pi \GN}{1+\Bar{f}} \int_{x'} {}_{(1)}\overline{G}^{+} [x-x'] \tau_{\mu\nu} [x'] . \label{eq:linDW2}
\end{eqnarray}
It is simpler than the linearized solutions in the first DW model. In fact, it is the GR solution divided by $1 + \bar{f}$:
\begin{eqnarray}
	\bar{h}_{\mu \nu}[\text{DWII}] = \frac{\bar{h}_{\mu\nu}[\text{GR}]}{1 + \bar{f}}. 
\end{eqnarray}
\textbf{Quadratic Order} \qquad We now turn to plugging the linearized solution $\bar{h}_{\mu\nu}$ into the second order terms of the nonlocal $\Delta G_{\mu\nu}$ in eq. \eqref{eq:energy}, so as to extract its contribution to the energy-momentum of the gravitational waves.
\begin{eqnarray}
	\delta_{2} \Delta G_{\mu\nu} 
	&=& \Bar{f}'  \left( \eta_{\mu\nu} \partial^{2} - \partial_{\mu} \partial_{\nu} \right) \left( -2 \int_{x'} {}_{(1)}\overline{G}^{+} \partial_{\mu'}  \left( \int_{x''} {}_{(1)}\overline{G}^{+} \delta_{1} R \ \ \partial^{\mu'} \int_{x'''} {}_{(1)}\overline{G}^{+} \delta_{1} R \right) \right. \notag \\
	&& \qquad \qquad \qquad \qquad \left. + \int_{x'} {}_{(1)}\overline{G}^{+} \left( \partial^{\mu'} \int_{x''} {}_{(1)}\overline{G}^{+} \delta_{1} R \ \ \partial_{\mu'} \int_{x'''} {}_{(1)}\overline{G}^{+} \delta_{1} R \right) \right)  \label{eq:secondeqom}
\end{eqnarray}
The first order Ricci scalar in the de Donder gauge is
\begin{eqnarray}
	\delta_{1} R 
	&=& \frac{1}{2} \partial^{2} \bar{h} 
	= - \frac{8 \pi \GN}{1+\Bar{f}} \tau ,
\end{eqnarray}
where in the second equality we have employed the trace of the equation of motion \eqref{eq:linDW2},
\begin{eqnarray}
	\partial^{2} \bar{h} =  - \frac{16 \pi \GN}{1+\Bar{f}} \tau . \label{eq:parttraceh}
\end{eqnarray}
The integral of the retarded Green function against the Ricci scalar yields
\begin{eqnarray}
\int {}_{(1)}\overline{G}^{+} \delta_{1} R 
\equiv \int_{x'} {}_{(1)}\overline{G}^{+}[x-x'] \delta_{1} R[x']
= \frac{1}{2} \bar{h}[x]. \label{eq:intricci}
\end{eqnarray}
Inserting eq. \eqref{eq:intricci} into \eqref{eq:secondeqom} now hands us
\begin{eqnarray}
	\delta_{2} \Delta G_{\mu\nu} 
	&=& \Bar{f}' \left(\eta_{\mu\nu} \partial^{2} - \partial_{\mu} \partial_{\nu} \right) \left(- \frac{1}{2}  \int_{x'} {}_{(1)}\overline{G}^{+} [x-x']  \bar{h} [x'] \left( - \frac{16 \pi \GN}{1+\Bar{f}} \tau [x'] \right) \right. \notag \\
	&& \ \ \ \ \ \ \ \ \ \ \ \ \ \ \ \ \ \ \ \ \ \ \  \left. - \frac{1}{4} \int_{x'} {}_{(1)}\overline{G}^{+} [x-x']  \partial_{\alpha'}  \bar{h} [x'] \partial^{\alpha'}\bar{h} [x']  \right), \label{eq:secondeqom2}
\end{eqnarray}
The gravitational wave is detected by the observer at $x^\mu = (t,\vec{x})$ in the far zone, where $r \equiv |\vec{x}| \to \infty$. If we further put $\vec{0}$ within the source of the GW, so that $r \gg r'$, then in the far zone,
\begin{eqnarray}
	|\vec{x}-\vec{x}'| = r-\vec{x}' \cdot \widehat{r} + \dots \simeq r ,
\end{eqnarray}
\noindent where $\widehat{r} \equiv \vec{x}/r$. 

To evaluate eq. \eqref{eq:secondeqom2} and similar integrals encountered in the ABN and VAAS models below, we use the approximation techniques from Poisson and Will \cite{gravity}, by breaking up the integrand into its contribution from the near and wave zone domains in 3D space. The boundary between these two domain is a sphere of arbitrary radius $\mathcal{R}$, where the near zone is $r' < \mathcal{R}$ and wave zone is $r' > \mathcal{R}$.

\begin{figure}[h!]
	\centering
	\includegraphics[scale=0.7]{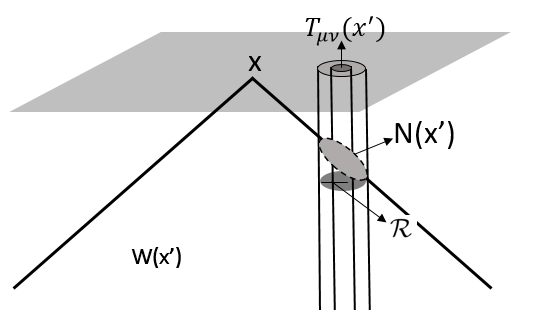}
	\caption{The integration domains is divided into near zone $N[x']$ and wave zones $W[x']$. The near zone area is the intersection between the past light cone of the observer at $x$ and the region immediately surrounding the world-tube of the source of GWs described by $T_{\mu\nu}[x']$; it is delineated by $r'< \mathcal{R}$. The wave zone $W[x']$, on the other hand, is the rest of the light cone, namely $r ' > \mathcal{R}$. }  \label{fig:nearwavezone}
\end{figure}

We begin with computing the near zone domain integration -- denoted with the subscript $\mathcal{N}$ -- which is the intersection between the past light cone of the field point $x$ and the near zone around the source. By inserting equations eq. \eqref{eq:green} and \eqref{eq:parttraceh} into eq. \eqref{eq:secondeqom2} gives
\begin{eqnarray}
	(\delta_{2} \Delta G_{\mu\nu})_{\mathcal{N}} &=& - \frac{1}{4 \pi} \left( \frac{ \GN}{1+\bar{f}} \right)^{2} \bar{f}' \left(\eta_{\mu\nu} \partial^{2} - \partial_{\mu} \partial_{\nu} \right) \left\{ 2\pi \int_{\Vec{x}'} \int_{\Vec{x}''}  \frac{\tau [t-|\vec{x}-\vec{x}'|,\Vec{x}'']}{|\Vec{x}-\Vec{x}'|}  \frac{\tau [t-|\vec{x}-\vec{x}'|,\Vec{x}']}{|\Vec{x}'-\Vec{x}''|}  \right. \notag \\
	&&  \ \ \ \ \ \ \ \ \ \ \ \ \ \left. + 4 \int_{\Vec{x}'} \int_{\Vec{x}''} \int_{\Vec{x}'''} \partial_{\alpha'}  \frac{\tau [t-|\vec{x}-\vec{x}'|,\Vec{x}'']}{|\Vec{x}'-\Vec{x}''|}  \partial^{\alpha'} \frac{\tau [t-|\vec{x}-\vec{x}'|,\Vec{x}''']}{|\Vec{x}'-\Vec{x}'''|}  \right\}.
\end{eqnarray}
We assume that the astrophysical source is non-relativistic, where the energy dominates over pressure density. More specifically, if $v \ll 1$ denotes the characteristic speed of the GW source's internal dynamics, then
\begin{align}
	\frac{T_{0i}}{T_{00}} \sim v
	\qquad \text{and} \qquad
	\frac{T_{ij}}{T_{00}} \sim v^2 .
\end{align}
At the quadratic order, $\tau \to T_{00} - T_{ii} \approx T_{00}$. It then follows that the near zone contribution of the nonlocal term to the GW stress tensor is
\begin{eqnarray}
	(t_{\mu\nu})_{\mathcal{N}} &=& \frac{\bar{f}' }{r} \frac{3 \GN}{8 \pi (1+\bar{f})^{2}}  \left( \delta^{0}_{\mu} - \widehat{r}^{i} \delta^{i}_{\mu}  \right) \left( \delta^{0}_{\nu} - \widehat{r}^{j} \delta^{j}_{\nu}  \right)  \partial^{2}_{0} \int_{\Vec{x}'} \int_{\Vec{x}''}  \frac{T_{00} [t-r,\Vec{x}'] T_{00} [t-r,\Vec{x}'']}{|\Vec{x}'-\Vec{x}''|} . \label{eq:resultnearzone}
\end{eqnarray}
This result is similar to that of the DWI model \cite{GWFluxDWI}, where the nonlocal contribution scales as $1/r$. The ensuing divergent GW flux can be avoided by setting $\bar{f}'$ to be zero.

Next, we examine the wave zone contributions, which we shall denote with the subscript $\mathcal{W}$. At second order in $\GN$, the $\tau$ in eq. \eqref{eq:secondeqom2} is simply $\eta^{\alpha\beta} \bar{T}_{\alpha\beta}$, the trace of the GW source's stress tensor. Hence, the first term of eq. \eqref{eq:secondeqom2} receives strictly near zone contributions, and what remains is
\begin{eqnarray}
	\left(\delta_{2} \Delta G_{\mu\nu}\right)_{\mathcal{W}} 
	&=& - \frac{\Bar{f}'}{4} \left(\eta_{\mu\nu} \partial^{2} - \partial_{\mu} \partial_{\nu} \right) \int_{x'} {}_{(1)}\overline{G}^{+} [x-x'] 
	\left(\partial_{\alpha'}  \bar{h} [x'] \partial^{\alpha'}\bar{h} [x']\right)_{\mathcal{W}}. \label{eq:wavezone}
\end{eqnarray}
Upon a non-relativistic expansion of the sources, we find that, at leading order in the non-relativistic expansion,
\begin{align}
	\left(\partial_{\alpha'}  \bar{h} \partial^{\alpha'}\bar{h}\right)_{\mathcal{W}}
	&\approx \mathcal{W}_\text{I} + \mathcal{W}_\text{II} ,\label{eq:r3} \\
\mathcal{W}_{1}
	&\equiv -\left(\frac{4 \GN}{1+\bar{f}} \right)^{2}  \frac{1}{r'^{4}} \left(M - P \right)^{2}, \label{psi1} \\ 
\mathcal{W}_{2}
	&\equiv \left(\frac{4 \GN}{1+\bar{f}} \right)^{2}  \frac{6 }{r'^{3}}  (M - P) \dot{P}; \label{psi2}
\end{align}
where $M$ and $P$ are the total mass and pressure,
\begin{eqnarray}
	M &\equiv& \int_{\vec{x}'} T_{00} [t-r, \vec{x}'] , \\
	P &\equiv& \int_{\vec{x}'} T_{mm} [t-r, \vec{x}']
\end{eqnarray}
and an overdot is a time derivative; namely, $\dot{P} \equiv \int_{\vec{x}'} \dot{T}_{mm}[t-r, \vec{x}']$. When eq. \eqref{eq:r3} is inserted into eq. \eqref{eq:wavezone}, the ensuing integrals take the form tackled by Poisson and Will \cite{gravity},
\begin{eqnarray}
	\label{WaveZoneIntegral_I}
	\psi_{\mathcal{W}}[x] \equiv \int_{\mathcal{W}} \frac{\mu[t- |\vec{x} - \vec{x}'|, \vec{x}']}{|\vec{x}-\vec{x}'|} \text{d}^{3} x' \label{wavezoneeq}. 
\end{eqnarray}
with the numerator restricted to the form
\begin{eqnarray}
	\label{WaveZoneIntegral_II}
	\mu[t',r',\theta',\phi'] = \frac{1}{4 \pi } \frac{F[t' - r']}{r'^{n}} n'^{\langle L \rangle} . \label{sourceWZ}
\end{eqnarray}
Here, $(\theta',\phi')$ are the angles on the $2-$sphere; $F$ is an arbitrary function of $t'-r'$; and $n^{\langle L \rangle}$ is a transverse and trace-free components of $\ell$ products of the unit radial vector $\widehat{r}[\theta',\phi']$. Then,
\begin{eqnarray}
	\psi_{\mathcal{W}} &=& \frac{n^{\langle L \rangle}}{r} \left(\int^{\mathcal{R}}_{0} \dd s F[t-r - 2s] A[s,r] + \int^{\infty}_{\mathcal{R}} \dd s F[t -r  - 2s] B[s,r] \right), \label{eq:wzPoisson}
\end{eqnarray}
where $s \equiv \frac{1}{2} (u-u') $, $u=t-r$; and, with $P_\ell$ denoting the $\ell$-th Legendre polynomial,
\begin{align}
\label{functionA}
A[s,r] 
&\equiv \int^{r+s}_{\mathcal{R}} \frac{P_{\ell}[\xi]}{r'^{n-1}} \dd r', \\
\label{functionB}
B[s,r] 
&\equiv \int^{r+s}_{s} \frac{P_{\ell}[\xi]}{r'^{n-1}} \dd r' , \\
\xi 
&\equiv 1+\frac{2s}{r}-2s\left(\frac{1}{r'}+\frac{s}{rr'}\right).
\end{align}
From equations \eqref{psi1} and \eqref{psi2}, we see that $(n,\ell)=(4,0)$ and $(n,\ell)=(3,0)$ for the cases at hand. In addition, as argued in \cite{gravity}, since the radius $\mathcal{R}$ is arbitrary; we expect it to cancel out at the end of the calculation. Taylor expanding in powers of $\mathcal{R}$ whenever appropriate, and keeping only $\mathcal{R}-$independent terms, we find the total contribution from the wave zone to be
\begin{eqnarray}
	\int_{x'} {}_{(1)}\overline{G}^{+} [x-x'] \left(\partial_{\alpha'}  \bar{h} [x'] \partial^{\alpha'}\bar{h} [x']\right)_{\mathcal{W}}
	&\approx&  \left(\frac{4 \GN}{1+\bar{f}} \right)^{2}  \left( \frac{  15 (M-P) \dot{P} }{  2   r }  + \frac{(M - P)^{2}}{2 r^{2}} \right).  
\end{eqnarray}
We arrive at the conclusion that the wave zone contribution to the GW energy-momentum, arising from the nonlocal gravitational interactions, also yields a $1/r$ piece that would produce a divergent flux at infinity:
\begin{eqnarray}
	(t_{\mu\nu})_{\mathcal{W}} 
	&=& - \frac{ \bar{f}' \GN} {4 \pi(1+\bar{f})^{2}}  \left( \delta^{0}_{\mu} - \widehat{r}^{i} \delta^{i}_{\mu}  \right) \left( \delta^{0}_{\nu} - \widehat{r}^{j} \delta^{j}_{\nu}  \right)   \partial^{2}_{0} \left( \frac{P \dot{P} }{r} + \dots \right). \label{wavez} \label{eq:resultfarzone}
\end{eqnarray}

\section{Amendola-Burzilla-Nersisyan (ABN)}
\label{Section_ABN}
The ABN model is derived from the studies of nonperturbative lattice quantum gravity \cite{ABN}. After varying the action in eq. \eqref{eq:ABN}, the corresponding nonlocal $\Delta G_{\mu\nu}$ in eq. \eqref{EoM} is revealed to be
\begin{eqnarray}
	\label{ABN_EoM}
	\Delta G_{\mu\nu} &=& - \frac{M^{4}}{6} \left(L R_{\mu\nu} - \nabla_{\mu} \nabla_{\nu} L + g_{\mu\nu} Q + \frac{1}{2} g_{\mu\nu} (S - R L) + \frac{1}{2} g_{\mu\nu}g^{\sigma\lambda} \left(\nabla_{\sigma} Q \nabla_{\lambda} S + \nabla_{\sigma} U \nabla_{\lambda} L \right) \right. \nonumber \\
	&& \left. - \frac{1}{2} \nabla_{\{\mu} U \nabla_{\nu\}} L - \frac{1}{2} \nabla_{\{\nu} Q \nabla_{\mu\}} S + \frac{1}{2} g_{\mu\nu} UQ \right).  \label{ABN_EoM_1}
\end{eqnarray}
The auxiliary fields obey the equations
\begin{eqnarray}
	\label{ABN_EoM_2}
	\Box U &=& -R, \ \ \ \ \ \ \ \ \ \ \ \ \Box Q = 1, \nonumber \\
	\Box S &=& U, \ \ \ \ \ \ \ \ \ \ \ \ \ \ \Box L = Q.
\end{eqnarray}
{\bf Minkowski Solution} \qquad We show that $g_{\mu\nu} = \eta_{\mu\nu}$ is a solution to ABN in vacuum, when $T_{\mu\nu}=0$. Firstly, flat spacetime means both the Einstein and Ricci tensors are zero; and so is the Ricci scalar. If we suppose $U, S, Q$ and $L$ depend only on time and not on space, their solutions to eq. \eqref{ABN_EoM_2} read
\begin{align}
	U = U_1 + U_2 t 
	\qquad \text{and} \qquad
	S = S_1 + S_2 t + U_1 \frac{t^2}{2} + U_2 \frac{t^3}{6},
\end{align}
where $U_{1,2}$ and $S_{1,2}$ are constants. Similarly, for constants $Q_{1,2}$ and $L_{1,2}$,
\begin{align}
	Q = \frac{t^2}{2} + Q_1 + Q_2 t 
	\qquad \text{and} \qquad
	L = \frac{t^4}{24} + L_1 + L_2 t + Q_1 \frac{t^2}{2} + Q_2 \frac{t^3}{6}.
\end{align}
(This assumption that $U,S,Q$ and $L$ depend solely on time is consistent with cosmological applications where $\gb_{\mu\nu} = a^2 \eta_{\mu\nu}$, with $a$ being the scale factor; Minkowski spacetime may be regarded as the $a \to 1$ limit.) Next, taking the trace of eq. \eqref{ABN_EoM_1} gives us
\begin{align}
\Delta G^\mu_{\phantom{\mu}\mu}	
&= -\left(\frac{1}{3} S_1 + \frac{1}{6} U_2 L_2 + \frac{1}{3} U_1 Q_1 + \frac{Q_1}{2} + \frac{1}{6} S_2 Q_2\right) M^4 \nonumber\\
	&\qquad\qquad
	-\frac{1}{2} \left(S_2+U_2 Q_1+\left(U_1+1\right) Q_2\right) M^4 t 
	-\frac{1}{4} \left(2 U_1+2 U_2 Q_2+1\right) M^4 t^2
	-\frac{1}{3} U_2 M^4 t^3 ;
\end{align}
which indicates, if we demand $\Delta G^{\mu}_{\phantom{\mu}\mu} = 0$,
\begin{align}
	\label{ABN_USQL_Constants_IofII}
	U_2 = 0, \qquad U_1 = -\frac{1}{2}, \qquad S_2 = -\sqrt{S_1+Q_1}, \qquad Q_2 = 2 \sqrt{S_1+Q_1} ;
\end{align} 
but leaves the rest of the constants unconstrained. Inserting these results back into eq. \eqref{ABN_EoM_1},
\begin{eqnarray}
	\Delta G^{0}_{\phantom{0}0}
	= -\frac{M^4}{8} (2S_1 + Q_1), \qquad
	\Delta G^{i}_{\phantom{i}j} = 0, \qquad
	\Delta G^{i}_{\phantom{i}j} 
	= \frac{M^{4}}{24} \delta^i_{\phantom{i}j} (2 S_1 + Q_1). \label{U1}
\end{eqnarray}
Hence, we see that
\begin{align}
	\label{ABN_USQL_Constants_IIofII}
	Q_1 = - 2S_1 
\end{align}
together with eq. \eqref{ABN_USQL_Constants_IofII} allows a Minkowski background solution
\begin{align}
	\gb_{\mu\nu} = \eta_{\mu\nu} .
\end{align}
For simplicity, throughout the rest of this section, we shall set all the integration constants in $U$, $S$, $Q$, and $L$ to zero except $U_1=-1/2$ in eq. \eqref{ABN_USQL_Constants_IofII}.

{\bf Setup and Auxiliary Fields} \qquad To study the ABN gravitational wave energy-momentum, we can write general spacetime metric and its small deviation
\begin{align}
	g_{\mu\nu} = \eta_{\mu\nu} + h_{\mu\nu} ,
	\qquad\qquad 
	|h_{\mu\nu}| \ll 1 ;
\end{align}
assume $h_{\mu\nu}$ is sourced by an isolated astrophysical system with stress tensor $T_{\mu\nu}$; and proceed to perturbatively solve eq. \eqref{EoM}, with the nonlocal $\Delta G_{\mu\nu}$ given by eq. \eqref{ABN_EoM}.

We begin by solving the auxiliary fields in eq. \eqref{ABN_EoM_2} up to second order in perturbations $\bar{h}_{\mu\nu}$. The first pair consists of $U = R/\Box$ and $S = U/\Box$. Keeping in mind $1/\Box = 1/\partial^2$ at zeroth order, as well as eq. \eqref{eq:intricci},
{\allowdisplaybreaks\begin{eqnarray}
		U &\approx& -\frac{1}{2} (1+\bar{h}) + \frac{1}{2} \int_{x'} \bar{h} \ {}_{(1)}\overline{G}^{+}  \delta_{1} R - \int_{x'} \delta_{1} {}_{(1)}\overline{G}^{+} \delta_{1} R - \int_{x'} {}_{(1)}\overline{G}^{+} \delta_{2} R , \\
		S &\approx& -\frac{1}{4} t^2 - \frac{1}{2} \int_{x'} {}_{(1)}\overline{G}^{+} \bar{h} + \frac{1}{2} \int_{x'} \bar{h} \ {}_{(2)}\overline{G}^{+}  \delta_{1} R - \int_{x'} \delta_{1} {}_{(2)}\overline{G}^{+} \delta_{1} R - \int_{x'} {}_{(2)}\overline{G}^{+} \delta_{2} R , \\
		Q 
		&\approx& \frac{t^2}{2} 
		+ \int_{x'} {}_{(1)}\overline{G}^{+} \left( \bar{h}^{0'0'} - \frac{1}{2} \bar{h}\right) \\
		&+& \int_{x'} {}_{(1)}\overline{G}^{+} \left( - \frac{3}{8} t' \partial_{0'} \bar{h}^{2}  + \frac{t'}{4} \partial_{t'} \left(  \bar{h}^{\rho' \sigma'}  \bar{h}_{\rho'\sigma'} \right) - t' \bar{h}^{\mu'}_{\phantom{\mu'} \sigma'} \partial_{\mu'}  \bar{h}^{\sigma' 0'}
		+ \frac{t'}{2} \partial_{\mu'} \left(  \bar{h}  \bar{h}^{\mu' 0'} \right)
		\right. \nonumber \\ 
		&& \left. \ \ \ \ \ \ \ \ \ \ \ \ \ \ \ \ \  - \bar{h}^{0'}_{\phantom{0'} \sigma'} \bar{h}^{\sigma' 0'} 
		+ \frac{1}{2} \bar{h} \bar{h}^{0'0'} 
		+ \frac{1}{2} \bar{h} t' \partial_{0'} \bar{h} 
		+ \bar{h}^{\mu'\nu'} \partial_{\mu'} \partial_{\nu'} \left\{ \int_{x''} {}_{(1)}\overline{G}^{+} \left( \bar{h}^{0''0''} - \frac{1}{2} \bar{h}\right) \right\}\right) \nonumber
\end{eqnarray}}
and
\begin{align}
	L
	&\approx \frac{t^4}{24}
	+ \int_{x'} {}_{(1)}\overline{G}^{+} t'^{2}   \left( \frac{1}{2} \bar{h}^{0'0'} - \frac{1}{4} \bar{h} \right) + \int_{x'} {}_{(2)}\overline{G}^{+}  \left( \bar{h}^{0'0'} - \frac{1}{2} \bar{h} \right) \nonumber\\
	&+ \int_{x'} {}_{(1)}\overline{G}^{+} \left(
	-  \frac{1}{16} t'^{3} \partial_{0'} \bar{h}^{2} + \frac{t'^{3}}{24} \partial_{0'} \left(  \bar{h}^{\rho' \sigma'}  \bar{h}_{\rho'\sigma'} \right)
	- \frac{t'^{3}}{6}  \bar{h}^{\mu'}_{\phantom{\mu'} \sigma'} \partial_{\mu'}  \bar{h}^{\sigma' 0'}
	+ \frac{t'^{3}}{12}  \partial_{\mu'} \left( \bar{h}  \bar{h}^{\mu' 0'} \right)
	\right. \nonumber  \\
	& \ \ \ \ \ \ \ \  \ \ \ \ \ \ \ \left. 
	- \frac{t'^{2}}{2} \bar{h}^{0'}_{\phantom{0'} \sigma'} \bar{h}^{\sigma' 0'}  
	+ \frac{t'^{2}}{4}  \bar{h} \bar{h}^{0'0'}
	+ \frac{t'^{3}}{24} \bar{h}  \partial_{0'} \bar{h}   
	\right. \nonumber \\
	&  \ \ \ \ \ \ \ \  \ \ \ \ \ \ \ \left. + \bar{h}^{\mu'\nu'} \partial_{\mu'} \partial_{\nu'} \left( \int_{x''} {}_{(1)}\overline{G}^{+} t''^{2}  \left( \frac{1}{2} \bar{h}^{0''0''} - \frac{1}{4} \bar{h} \right) +  \int_{x''} {}_{(2)}\overline{G}^{+}  \left( \bar{h}^{0''0''} - \frac{1}{2} \bar{h} \right) \right)  \right. \nonumber \\
	&  \ \ \ \ \ \ \ \  \ \ \ \ \ \ \ \left. 
	- \frac{1}{2}  \bar{h} \int_{x''} {}_{(1)}\overline{G}^{+} \left( \bar{h}^{0''0''} - \frac{1}{2} \bar{h}\right) 
	\right) \nonumber\\
	& + \int_{x'} {}_{(2)}\overline{G}^{+} \left(
	- \frac{3}{8} t' \partial_{0'} \bar{h}^{2}  + \frac{t'}{4} \partial_{0'} \left(  \bar{h}^{\rho' \sigma'}  \bar{h}_{\rho'\sigma'} \right) -  t' \bar{h}^{\mu'}_{\phantom{\mu'} \sigma'} \partial_{\mu'}  \bar{h}^{\sigma' 0'}
	+ \frac{t'}{2} \partial_{\mu'} \left(  \bar{h}  \bar{h}^{\mu' 0'} \right)
	\right. \nonumber \\ 
	& \left. \ \ \ \ \ \ \ \ \ \ \ \ \ \ \ \ \  
	- \bar{h}^{0'}_{\phantom{0'} \sigma'} \bar{h}^{\sigma' 0'} 
	+ \frac{1}{2} \bar{h} \bar{h}^{0'0'} 
	+ \frac{1}{2} \bar{h} t' \partial_{0'} \bar{h} 
	+ \bar{h}^{\mu'\nu'} \partial_{\mu'} \partial_{\nu'} \int_{x''} {}_{(1)}\overline{G}^{+} \left( \bar{h}^{0''0''} - \frac{1}{2} \bar{h}\right)\right) . 
\end{align}
We have introduced another retarded Green function,
\begin{align}
	{}_{(2)}\overline{G}^{+} [x-x'] = \Theta [t-t'-|\vec{x}-\vec{x}'|] ,
\end{align}
where $\Theta$ is the step function ($\Theta[x>0]=1$ and $\Theta[x<0]=0$). It obeys $\partial^4 {}_{(2)}\overline{G}^{+} = 8\pi \delta^{(4)}[x-x']$ and is related to ${}_{(1)}\overline{G}^{+} $ via the integral
\begin{align}
	{}_{(2)}\overline{G}^{+} [x-x''] = 8\pi \int_{x'}  {}_{(1)}\overline{G}^{+} [x-x'] {}_{(1)}\overline{G}^{+} [x'-x'']  .
\end{align}
{\bf Linear Dynamics} \qquad Next, we examine the ABN equations containing exactly one power of the metric perturbation. Using the de Donder gauge linearized Einstein tensor in eq. \eqref{Einstein_1stOrder_deDonder}, the ABN equations become
\begin{align}
	\label{ABN_Linearized_EoM}
	-\frac{1}{2} \partial^2 \bar{h}_{\mu\nu} + \delta_1 \Delta G_{\mu \nu}[h]
	&= 8\pi \GN \bar{T}_{\mu \nu} .
\end{align}
The integral version of eq. \eqref{ABN_Linearized_EoM} is
\begin{eqnarray}
	\bar{h}_{\mu \nu} 
	= -16 \pi \GN \int_{x'} {}_{(1)}\overline{G}^{+} \bar{T}_{\mu'\nu'} + 2 \int_{x'} {}_{(1)}\overline{G}^{+} \delta_{1} \Delta G_{\mu'\nu'} + \mathcal{O} [M^{8}] .\label{linearizedABN}
\end{eqnarray}
Because $\delta_{1} \Delta G_{\mu \nu}$ is of order $M^4$, we see that the zeroth order $M$-independent solution of the metric perturbation is simply that of the linearized Einstein's equation:
\begin{align}
	\bar{h}_{\mu\nu}[M^0] 
	= -16 \pi \GN \int_{x'} {}_{(1)}\overline{G}^{+} \bar{T}_{\mu\nu}
	= -4 \GN \int_{\mathbb{R}^3} \dd^3 \vec{x}' \frac{\bar{T}_{\mu\nu}[t-|\vec{x}-\vec{x}'|,\vec{x}']}{|\vec{x}-\vec{x}'|} . \label{Einsteinlinearized}
\end{align}
Similar to the Born series, each higher order in $M^4$ solution to eq. \eqref{linearizedABN} can be obtained by repeated iteration. The solution accurate up to $M^4$ order is simply eq. \eqref{linearizedABN} but with eq. \eqref{Einsteinlinearized} inserted into
\begin{eqnarray}
	\delta_{1} \Delta G_{\mu\nu}
	&=&   -  \frac{M^{4}}{6} \left(-
	\frac{t^4}{48}  \partial^{2}  \bar{h}_{\mu\nu}  - \partial_{\mu} \partial_{\nu}  \left( \int_{x'} {}_{(1)}\overline{G}^{+} t'^{2}  \left( \frac{1}{2} \bar{h}^{0'0'} - \frac{1}{4} \bar{h} \right) + \int_{x'} {}_{(2)}\overline{G}^{+}  \left( \bar{h}^{0'0'} - \frac{1}{2} \bar{h} \right) \right) \right. \nonumber \\ 
	&& \qquad \qquad \left. + \frac{t^3}{12} \left(\partial_{\mu}  \bar{h}_{\nu 0} + \partial_{\nu} \bar{h}_{\mu 0}  - \partial_{0}  \bar{h}_{\mu\nu} \right)   
	- \frac{1}{4} t^2 \eta_{\mu\nu}  \bar{h}^{00} 
	-  \frac{t^4}{96} \eta_{\mu \nu} \partial ^{2} \bar{h}   \right. \nonumber  \\
	&& \qquad \qquad \left. + \frac{1}{2} \eta_{\mu\nu} t \partial_{0} \int_{x'} {}_{(1)}\overline{G}^{+} \bar{h}^{0'0'}  - \frac{1}{2} \eta_{\mu\nu} t \partial_{0} \int_{x'} {}_{(1)}\overline{G}^{+}  \bar{h}  + \frac{t}{8} \delta^{0}_{\{\nu} \partial_{\mu\}} \int_{x'} {}_{(1)}\overline{G}^{+} \bar{h}    \right. \nonumber \\
	&& \qquad \qquad \left.  + \frac{t}{4} \delta^{0}_{\{\nu} \partial_{\mu\}}  \int_{x'} {}_{(1)}\overline{G}^{+} \bar{h}^{0'0'}  
	- \frac{5}{8} \eta_{\mu\nu}  \int_{x'} {}_{(1)}\overline{G}^{+}  \bar{h}
	+ \frac{3}{4} \eta_{\mu\nu}  \int_{x'} {}_{(1)}\overline{G}^{+}  \bar{h}^{0'0'} 
	\right)  . \label{firstABN} 
\end{eqnarray}
{\bf Quadratic Order} \qquad This $\mathcal{O}[\GN M^4]$ accurate solution derived from eq. \eqref{linearizedABN} is then plugged into $\delta_2 \Delta G_{\mu\nu}$. By organizing the result in powers of observer-source distance $r$, we find that the dominant contribution for large $r$ arises from
\begin{eqnarray}
	\delta_{2} \Delta G_{\mu\nu} &\approx& \frac{1}{6} M^{4} \partial_{\mu} \partial_{\nu} \delta_{2} L + \dots,
\end{eqnarray}
where the second order piece of $L$ is 
\begin{eqnarray}
	\label{eq:delta2L} 
	\delta_{2} L &=& \int_{x'} {}_{(1)}\overline{G}^{+} \left(\frac{1}{4} \bar{h}^{2} t'^{2} - \frac{1}{12} \bar{h}^{\mu'0'} t'^{3} \partial_{\mu'} \bar{h } + \frac{1}{12} \bar{h} t'^{3} \partial_{0'} \bar{h} - \frac{1}{4} \bar{h} \bar{h}^{0'0'} t'^{2}  - \frac{1}{2}  \bar{h} \int_{x''} {}_{(1)}\overline{G}^{+} \left( \bar{h}^{0''0''} - \frac{1}{2} \bar{h}\right)  \right. \notag \\
	&& \ \ \ \ \ \ \ \ \ \  \ \ \ \ \ \ \ \left. + \bar{h}^{\mu'\nu'} \partial_{\mu'} \partial_{\nu'} \left( \int_{x''} {}_{(1)}\overline{G}^{+} t''^{2}  \left( \frac{1}{2} \bar{h}^{0''0''} - \frac{1}{4} \bar{h} \right) +  \int_{x''} {}_{(2)}\overline{G}^{+}  \left( \bar{h}^{0''0''} - \frac{1}{2} \bar{h} \right) \right) \right)   \\
	&& + \int_{x'} {}_{(2)}\overline{G}^{+} \left(\frac{1}{2} \bar{h}^{2} - \frac{1}{2} \bar{h}^{\mu'0'} t' \partial_{\mu'} \bar{h} + \frac{1}{2} \bar{h}t' \partial_{0'} \bar{h} - \frac{1}{2} \bar{h} \bar{h}^{0'0'} + \bar{h}^{\mu'\nu'} \partial_{\mu'} \partial_{\nu'} \int_{x''} {}_{(1)}\overline{G}^{+} \left( \bar{h}^{0''0''} - \frac{1}{2} \bar{h}\right)\right). \notag
\end{eqnarray}
We find the near zone contribution from eq. \eqref{eq:delta2L} to $(-8\pi\GN)^{-1} \delta_2 \Delta G_{\mu\nu} \subset t_{\mu\nu}$ to scale at least as $r^2$:
\begin{align}
	\label{ABN_GWStressTensor_N}
	&(t_{\mu \nu })_{\mathcal{N}}
	=  \frac{ M^{4}  \GN}{32\pi } r^{2}   \left( \delta^{0}_{\mu} -  \delta_{i}^{\mu} \widehat{r}^{i} \right) \left( \delta^{0}_{\nu} -  \delta^{j}_{\nu} \widehat{r}^{j} \right)  \\
	& \times \partial^{2}_{t} \Bigg\{ (t - r)  M   \int_{\vec{y}'} \partial_{t} \bar{T}_{ll}[t-r, \vec{y}'] 
	+ \widehat{r}^{i} \widehat{r}^{j}   \int_{\vec{y}} \partial^{2}_{t} \bar{T}^{ij} [t-r , \vec{y}]  \int^{t-r}_{- \infty}  \text{d} y'^{0} y'^{0} \left( M + \int_{\vec{y}'} \bar{T}_{ll}[y'^{0}, \vec{y}'] \right)  \nonumber \\
	&\qquad + 2 \widehat{r}^{i} \widehat{r}^{j}  (t-r)  \int_{\vec{y}} \partial_{t}  \bar{T}^{ij} [t-r , \vec{y}] \left( M  +   \int_{\vec{y}'} \partial_{t} \bar{T}_{ll}[y'^{0}  , \vec{y}'] \right) - \widehat{r}^{i} \widehat{r}^{j}   \int_{\vec{y}} \bar{T}^{ij} [t-r , \vec{y}] \int_{\vec{y}'}  \bar{T}_{ll}[t-r, \vec{y}']  \Bigg\} + \dots.  \nonumber
\end{align}
Next, we evaluate the wave zone contribution from eq. \eqref{eq:delta2L}.
\begin{align}
	&(\delta_{2} \Delta G_{\mu\nu})_{\mathcal{W}} \nonumber\\
	\label{ABN_delta2DeltaG_W}
	&\approx \frac{1}{12} M^{4} \GN^{2} \partial_{\mu} \partial_{\nu} \partial_{\alpha} \partial_{\beta} \int_{x'} {}_{(1)}\overline{G}^{+} [x-x'] \frac{1}{  |\vec{x}'| } \int_{\vec{y}} \bar{T}^{\alpha \beta } [t' - |\vec{x}'|, \vec{y}] \\
	&\times \Bigg\{ \left(   \int^{t'  - |\vec{x}'|}_{- \infty} \text{d} y'^{0}  (t'+y'^{0})^{2} \left( M + \int_{\vec{y}'} \bar{T}_{ll}[y'^{0}, \vec{y}'] \right) \right. 
	\left. + \frac{|\vec{x}'|^{2} }{3} \int^{t'  - |\vec{x}'|}_{- \infty} \text{d} y'^{0}   \left( M + \int_{\vec{y}'} \bar{T}_{ll}[y'^{0}, \vec{y}'] \right)  \right) \nonumber \\
	&\qquad  + \left(  \int^{t'  - |\vec{x}'|}_{- \infty} \text{d} y'^{0}  (t'-y'^{0})^{2} 
	\left( M + \int_{\vec{y}'} \bar{T}_{ll}[y'^{0}, \vec{y}'] \right) \right. 
	\left. +|\vec{x}'|^{2}  \int^{t'  - |\vec{x}'|}_{- \infty} \text{d} y'^{0} \left( M + \int_{\vec{y}'} \bar{T}_{ll}[y'^{0}, \vec{y}'] \right)  \right) \Bigg\} +\dots. \nonumber
\end{align}
Comparing eq. \eqref{ABN_delta2DeltaG_W} with equations \eqref{wavezoneeq} and \eqref{sourceWZ}, we may identify
\begin{eqnarray}
	\label{ABN_mu}
	\mu^{\alpha \beta} [t', r'] 
	&= \frac{1}{4\pi r'} \cdot f_{(1)}^{\alpha\beta}[t'-r'] + \frac{r'}{4\pi} \cdot f_{(-1)}^{\alpha\beta}[t'-r'] ;
\end{eqnarray}
where the first term on the right hand side of eq. \eqref{ABN_mu} corresponds to $n=1$, $n'^{\langle L \rangle}=1$, $\ell=0$, and
\begin{align}
	\label{ABN_fp1}
	f_{(1)}^{\alpha\beta}[t'-r']
	&= \frac{1}{16(4 \pi)^{2}} \int_{\vec{y}} \bar{T}^{\alpha \beta } [t' - r', \vec{y}] 
	\int^{t'-r'}_{-\infty} \dd y'^{0} \left( (t'+y'^{0})^{2} + \frac{1}{2} (t'-y'^{0})^{2} \right) \left( M + \int_{\vec{y}'} \bar{T}_{ll}[y'^{0}, \vec{y}'] \right) ; \nonumber
\end{align}
whereas the second term corresponds to $n=-1$, $n'^{\langle L \rangle}=1$, $\ell=0$, and
\begin{align}
	f_{(-1)}^{\alpha\beta}[t'-r']
&= \frac{5}{96(4 \pi)^{2}} \int_{\vec{y}} \bar{T}^{\alpha \beta } [t' - r', \vec{y}] \int^{t'- r'}_{- \infty} \dd y'^{0} \left( M + \int_{\vec{y}'} \bar{T}_{ll}[y'^{0}, \vec{y}'] \right).
\end{align}
We have regulated the time integrals by assuming the matter source has been turned on only at $t=0$. The wave zone contribution to $(-8\pi\GN)^{-1} \delta_2 \Delta G_{\mu\nu} \subset t_{\mu\nu}$ also turns out to scale as $r^2$:
\begin{align}
\label{ABN_GWStressTensor_W}
&(t_{\mu \nu})_{\mathcal{W}}
= - \frac{5 M^{4} \GN}{864 \pi} r^{2} \left( \delta^{0}_{\mu} -  \delta_{i}^{\mu} \widehat{r}^{i} \right) \left( \delta^{0}_{\nu} -  \delta^{j}_{\nu} \widehat{r}^{j} \right)  \\
& \times \partial^{2}_{t} \left\{ \frac{1}{2}  M \left( \int_{\vec{y}} \bar{T}_{ll}[t-r, \vec{y}] 
	+ \widehat{r}^{i} \widehat{r}^{j} \int_{\vec{y}} \bar{T}^{ij} [t-r, \vec{y}] 
	+ \widehat{r}^{i} \widehat{r}^{j} (t-r)  \int_{\vec{y}} \partial_{t} \bar{T}^{ij} [t-r, \vec{y}]  \right) \right. \nonumber \\
& \qquad \left. + \frac{1}{2} \int_{\vec{y}} \partial_{t} \bar{T}^{ij} [t-r, \vec{y}]  \int^{t-r}_{- \infty} \text{d} y'^{0} \int_{\vec{y}'} \bar{T}_{ll}[y'^{0}, \vec{y}']  \right. 
	\left. +  \widehat{r}^{i} \widehat{r}^{j} 
	\int^{\infty}_{\mathcal{R}} ds \int_{\vec{y}} \partial_{t} \bar{T}^{ij} [t-r-2s, \vec{y}] \int_{\vec{y}'} \bar{T}_{ll}[t-r-2s, \vec{y}'] \right\}. \nonumber
\end{align}
Finally, we examine the nonlocal GW energy-momentum contributions due to the second order Einstein tensor $\delta_2 G_{\mu\nu}$ in eq. \eqref{eq:energy}. (This issue is not relevant for the DWII model because its linearized solutions did not involve nonlocal interactions.) Since $\delta_2 G_{\mu\nu}$ itself consists of spacetime-local products of the metric perturbation $\bar{h}_{\mu\nu}$ and its derivatives -- schematically, $(\partial \bar{h})^2$ or $\bar{h}\partial^2\bar{h}$ -- at lowest order in $M$ we merely need to consider the cross term between the GR linear solution with the $\mathcal{O}[\GN M^4]$ one we sketched above. A typical term reads
\begin{align}
	\label{ABN_GWStressTensor_Einstein}
	\bar{h}^{ij}[\text{GR}] \partial_{0} \partial_{k} \bar{h}_{ij}[\GN M^{4}]  
	&\approx - \frac{1}{36} M^{4}  \GN^{2} \left( \frac{(2t-r)^{3}}{r} + (2t-r) r\right) \widehat{r}^{k} \nonumber\\
	&\qquad\qquad
	\times \int_{\vec{x}'} \bar{T}^{ij} [t-r, \vec{x}'] \int_{\vec{x}''}
	\ddot{\bar{T}}_{ij} [t-r, \vec{x}''] + \dots . 
\end{align}
To sum, both the second order nonlocal $\delta_2 \Delta G_{\mu\nu}$ and the Einstein tensor $\delta_2 G_{\mu\nu}$ produce a GW energy-momentum at first order in nonlocality (and, hence, in $M^{4}$) that goes at least as $r^{2}$. Hence, the far zone GW flux is divergent.

\section{Vardanyan-Akrami-Amendola-Silvestri (VAAS)}
\label{Section_VAAS}
The construction of the VAAS model originated from bimetric theories of gravity \cite{VAAS}. The simplest nonlocal bimetric model consists of two metrics that interact nonlocally through the Ricci scalar. Varying the action in eq. \eqref{eq:VAAS}, the nonlocal term in eq. \eqref{EoM} is revealed to be
\begin{eqnarray}	
	\label{VAAS_EoM_1}	
	\Delta G_{\mu\nu}
	&\equiv& -2 \alpha V G_{\mu\nu} - m^2 \left(1-\frac{U}{2}\right) g_{\mu\nu}
	+ 2\alpha \nabla_\mu \nabla_\nu V + \alpha \nabla^\rho U \nabla_\rho V g_{\mu\nu} \nonumber\\
	&&- \alpha \left( \nabla_\mu U \nabla_\nu V + \nabla_\nu U \nabla_\mu V \right) , 
\end{eqnarray}
where the auxiliary fields are defined by
\begin{align}	
	\label{VAAS_EoM_2}
	\Box U &= R, \qquad\qquad \Box V = \frac{m^2}{2\alpha} .
\end{align}
{\bf No Minkowski Solution} \qquad We first show that $g_{\mu\nu} = \eta_{\mu\nu}$ cannot be a solution to VAAS in vacuum, i.e., when $T_{\mu\nu}=0$. Suppose flat spacetime were a vacuum solution, then the Einstein and Ricci tensors are zero; and so is the Ricci scalar. If we further suppose $V$ and $U$ depend only on time and not on space, their solutions to eq. \eqref{VAAS_EoM_2} read
\begin{align}
	U = U_1 + U_2 t 
	\qquad \text{and} \qquad
	V = V_1 + V_2 t + \frac{m^2}{4\alpha} t^2 ,
\end{align}
where $U_{1,2}$ and $V_{1,2}$ are constants. (Just like the ABN model, this assumption that $V$ and $U$ depend solely on time is consistent with cosmological applications; Minkowski spacetime may be regarded as the $a \to 1$ limit.) Next, taking the trace of eq. \eqref{EoM} hands us
\begin{align}
	m^2(-3 + 2 U_1) + 2 \alpha U_2 V_2 + 3 m^2 U_2 t = 0 .
\end{align}
Setting to zero both the constant term and the coefficient of $t$ indicates $U$ is  constant but leaves $V_{1,2}$ unconstrained. Inserting these results back into eq. \eqref{EoM}, we arrive at
\begin{align}
	m^2 \delta_\mu^0 \delta_\nu^0 = m^2 \left(1-\frac{U}{2}\right)\eta_{\mu\nu}.
\end{align}
If $m \neq 0$, $U$ is fixed by the $\mu\nu=00$ equation; but the spatial $\mu\nu=ij$ components cannot be satisfied because the left hand side is zero. The only consistent solution is therefore $m=0$, when the nonlocal terms are absent.

{\bf Vacuum Cosmology in Fermi-Normal-Coordinates (FNC)} \qquad Since empty flat spacetime does not exist in VAAS, we shall turn to its vacuum cosmological counterpart. Moreover, it does not appear easy to obtain an exact analytic vacuum solution to VAAS. To make further progress, we shall therefore resort to employing a Fermi-Normal-Coordinate (FNC) system, so that we may not only obtain such a vacuum cosmological solution perturbatively; flat spacetime techniques may continue to be used to perform analytic calculations. To this end, we shall assume the nonlocal mass parameter $m$ is such that $1/m$ characterizes the size of the observable universe, and is therefore the largest length scale in what follows. In particular, the observer-source distance is merely of astrophysical scales and much smaller than $1/m$. Now, it is known that spatially flat FLRW universes, described by the line element
\begin{align}
	\gb_{\mu\nu} \dd x^\mu \dd x^\nu = \dd t^2 - a[t]^2 \dd\vec{x} \cdot \dd\vec{x} ,
\end{align}
may always at leading order in the Hubble parameter be put in the FNC form
\begin{align}
	\label{FNC}
	\gb_{\mu\nu} \dd x^\mu \dd x^\nu
	= \left( 1 + \varphi[t] (\Delta \vec{X})^2 \right) \dd t^2 - \left( 1 + \Pi[t] (\Delta \vec{X})^2 \right) \dd\vec{x}^2 ,
\end{align}
where $\Delta \vec{x} \equiv \vec{x} - \vec{x}_0$ is the coordinate spatial displacement from the free falling `observer' at $\vec{x}_0$; whereas $\varphi$ and $\Pi$ only depend on time and not on space. In a vacuum cosmological geometry, the only dimension-ful parameter available is $m$, and we may therefore deduce by dimensional analysis alone that
\begin{align}
	\varphi \sim \Pi \sim m^2 (m \cdot t)^p ,
\end{align}
where $p$ is some positive power. (In fact, we shall see, $p=0$.) At leading order in $m$, eq. \eqref{VAAS_EoM_2} reads
\begin{align}
	\label{VAAS_VacuumCosmology_EoM_UV}
	\ddot{U} \approx 3 (4\Pi + 2 \varphi - \Delta\vec{X}^2 \ddot{\Pi})
	\qquad \text{and} \qquad
	\ddot{V} \approx \frac{m^2}{2\alpha} ,
\end{align}
where each overdot denotes a derivative with respect to $t$; and therefore both $U$ and $V$ must scale as the second power of $m$ or higher. Specifically, we have
\begin{align}
	\label{VAAS_VacuumCosmology_VSoln}
	V[t] = \frac{m^2 t^2}{4 \alpha} + \dots ,
\end{align}
where we have chosen to set the integration constants to zero because we want to recover flat spacetime -- i.e., the vacuum solution of General Relativity -- when $m=0$ and the nonlocal terms are discarded. With all these scaling relations in mind, we develop the Einstein tensor to leading order in $m$ and find
\begin{align}
	\label{VAAS_BackgroundEinstein00}
	G_{00} &\approx - 6 \Pi, \\
	\label{VAAS_BackgroundEinstein0i}
	G_{0i} &\approx - 2 \Delta X^i \dot{\Pi}, \\
	\label{VAAS_BackgroundEinsteinij}
	G_{ij} &\approx \delta_{ij} \left( 2 (\Pi + \varphi) - \Delta\vec{X}^2 \ddot{\Pi} \right) .
\end{align}
Similarly, by discarding the $V G_{\mu\nu}$, $m^2 (U/2) g_{\mu\nu}$, and $\nabla U \nabla V$ terms, which scale at least as four powers of $m$, we infer eq. \eqref{VAAS_EoM_1} reduces to
\begin{align}
	\label{VAAS_BackgroundDeltaGmunu}
	-\Delta G_{\mu\nu} &\approx m^2 (\eta_{\mu\nu} - \delta_\mu^0 \delta_\nu^0) .
\end{align}
Setting $G_{\mu\nu} = -\Delta G_{\mu\nu}$, we find that a vacuum cosmology in the VAAS model at first order in the FNC expansion admits the following solutions:
\begin{align}
	\Pi = 0 \qquad \text{and} \qquad \varphi = -\frac{m^2}{2} .
\end{align}
Incidentally, the $\Pi=0$ renders the $U$ equation in eq. \eqref{VAAS_VacuumCosmology_EoM_UV} consistent; because the $-\Delta\vec{X}^2 \ddot{\Pi}$ term drops out and $\ddot{U}$ then depends only on time and not on space
\begin{align}
	\label{VAAS_VacuumCosmology}
	\gb_{\mu\nu} \dd x^\mu \dd x^\nu
	= \left( 1 - \frac{1}{2} (m \cdot \Delta\vec{X})^2 \right) \dd t^2 - \dd\vec{x}^2 + \mathcal{O}\left[\left(m \Delta\vec{X}\right)^4\right].
\end{align}
Note that eq. \eqref{VAAS_BackgroundDeltaGmunu} also transparently demonstrates, de Sitter spacetime itself cannot be a solution of the VAAS model because it would have to satisfy $G_{\mu\nu} = - \Delta G_{\mu\nu} = \Lambda \eta_{\mu\nu}$ in a locally flat frame, for some appropriate positive constant $\Lambda$.

\noindent {\bf Setup and Auxiliary Fields} \qquad We now turn to studying gravitational waves propagating on top of the `background' vacuum cosmological spacetime in eq. \eqref{VAAS_VacuumCosmology}; namely, we seek to expand and proceed to solve eq. \eqref{EoM} perturbatively
\begin{align}
	g_{\mu\nu} 
	&= \gb_{\mu\nu} + h_{\mu\nu} \\
	&\approx \eta_{\mu\nu} - \frac{1}{2} (m \Delta \vec{X})^2 \delta_\mu^0 \delta_\nu^0 + h_{\mu\nu}
\end{align}
to first order in $m^2$ but second order in $h_{\mu\nu}$. But since the order $m^2$ contributions to the auxiliary fields solve the `background' vacuum cosmology equations-of-motion we have just obtained, what remains are the first and second order in $h_{\mu\nu}$ terms. The $\Box U = R$ in eq. \eqref{VAAS_EoM_2}, for instance, now reads
\begin{align}
	\frac{1}{\sqrt{|g|}} \partial_\mu \left( \sqrt{|g|} g^{\mu\nu} \partial_\nu \left(\delta_1 U + \delta_2 U\right) \right)
	= \delta_1 R + \delta_2 R ;
\end{align}
except now $g_{\mu\nu} = \eta_{\mu\nu} + h_{\mu\nu}$. Turning this into an integral equation for each order in the perturbation,
\begin{eqnarray}
	\delta_1 U 
	&=& - \frac{1}{2} h , \\
	\delta_2 U 
	&=& \frac{1}{2} \int_{x'} h \ {}_{(1)}\overline{G}^{+}  \delta_{1} R + \int_{x'} \delta_{1} {}_{(1)}\overline{G}^{+} \delta_{1} R + \int_{x'} {}_{(1)}\overline{G}^{+} \delta_{2} R .
\end{eqnarray}
Similar lines of analysis will lead us to the following integral-equations-of-motion for $V$.
\begin{eqnarray}
	\label{VAAS_delta1V}
	\delta_{1} V
	&=& \int_{x'} {}_{(1)}\overline{G}^{+} \left(  \frac{1}{2}  \bar{h}^{0'0'}  - \frac{1}{4} \bar{h}\right) , \\
	\label{VAAS_delta2V}
	\delta_{2} V 
	&=& \int_{x'} {}_{(1)}\overline{G}^{+} \left(\frac{1}{4} \bar{h}^{2} - \frac{1}{4} \bar{h}^{\mu'0'} t' \partial_{\mu'} \bar{h} + \frac{1}{4} \bar{h} t' \partial_{0'} \bar{h} - \frac{1}{4} \bar{h} \bar{h}^{0'0'} + \bar{h}^{\mu'\nu'} \partial_{\mu'} \partial_{\nu'} \delta_{1} V \right) .
\end{eqnarray}
{\bf Linear Dynamics} \qquad We now turn to the gravitational dynamics at linear order in metric perturbations:
	\begin{align}
		\label{VAAS_Linearized_EoM}
		\delta_{1} G_{\mu\nu}[h] + \delta_1 \Delta G_{\mu\nu} 
		= 8\pi \bar{T}_{\mu\nu} ,
	\end{align}
	where $\bar{T}_{\mu\nu}$ is the matter stress tensor evaluated on $\gb_{\mu\nu}$ in eq. \eqref{VAAS_VacuumCosmology}; while the linearized Einstein tensor is
\begin{align}
	\delta_{1} G_{\mu\nu}[h]  &=  \frac{1}{2} \left( - \overline{\Box} h_{\mu\nu} + \overline{\nabla}_{\alpha} \overline{\nabla}_{\mu} h_{\nu}^{\phantom{\nu}\alpha} + \overline{\nabla}_{\alpha} \overline{\nabla}_{\nu} h_{\mu}^{\phantom{\mu}\alpha} - \overline{g}_{\mu\nu}  \overline{\nabla}_{\alpha} \overline{\nabla}_{\beta} h^{\alpha \beta} + \gb_{\mu \nu} \overline{\Box} h -  \overline{\nabla}_{\nu}  \overline{\nabla}_{\mu} h \right) ,
\end{align}
where all the covariant derivatives are with respect to $\gb_{\mu\nu}$ in eq. \eqref{VAAS_VacuumCosmology}.

By expanding eq. \eqref{VAAS_Linearized_EoM} up to first order in $m^2$, we will thus capture the leading modifications of this linearized GR solution arising from the presence of the nonlocal dynamics encoded in $\Delta G_{\mu\nu}$. Moreover, the integral form of eq. \eqref{VAAS_Linearized_EoM} up to order $m^2$ is
\begin{eqnarray}
	\bar{h}_{\mu\nu} = -16 \pi \GN \int_{x'} {}_{(1)}\overline{G}^{+} \bar{T}_{\mu'\nu'}+ 2 \int_{x'} {}_{(1)}\overline{G}^{+} \delta_{1} G_{\mu'\nu'} [m^{2}] + 2 \int_{x'} {}_{(1)}\overline{G}^{+} \delta_{1} \Delta G_{\mu'\nu'} + \mathcal{O} [m^{4}] ; \label{VAAS_hbar}
\end{eqnarray}
where
\begin{eqnarray}
	\delta_{1} \Delta G_{\mu\nu}[m^2] 
	&\equiv& m^{2} \left( \frac{1}{4} t^{2} \partial^{2} \bar{h}_{\mu\nu}  - \frac{1}{2} t \partial_{\{\mu}  \bar{h}_{\nu\}0}  + \frac{1}{2} t \partial_{0} \bar{h}_{\mu\nu} + \frac{3}{4}\eta_{\mu\nu}  \bar{h} -   \bar{h}_{\mu\nu} \right.  \nonumber \\
	&&	 \left.  \qquad
	+   \partial_{\mu} \partial_{\nu}  \int_{x'} {}_{(1)}\overline{G}^{+} \left(   \bar{h}^{0'0'}[x']  - \frac{1}{2} \bar{h}[x'] \right)  \right) ;\label{firstVAAS}
\end{eqnarray}
and the linearized Einstein tensor, written in an FNC expansion, takes the form
\begin{eqnarray}
	\delta_{1} G_{\mu \nu} &=& - \frac{1}{2} \partial^{2} \bar{h}_{\mu \nu} + \frac{3}{2} m^{2} \bar{h}_{\mu \nu} - \frac{1}{2} \eta_{\mu \nu} m^{2} \left( \frac{1}{2} \bar{h} + \bar{h}^{00} \right) \nonumber \\
	&& + \frac{1}{8} m^{2} \eta_{\mu \nu} \left(\bar{h} - 2 \bar{h}^{0}_{\phantom{0} 0} + 2 \Delta X^{a} \partial_{a} \bar{h} - 2 \Delta X^{a} \partial_{a}  \bar{h}^{0}_{\phantom{0} 0} \right) \nonumber \\
	&&  - \frac{1}{4} m^{2} \left(  2 \bar{h}_{\nu0} \delta^{0}_{\mu} + 2 \bar{h}_{\mu 0} \delta^{0}_{\nu} +  \bar{h}  \delta^{0}_{\mu } \delta^{0}_{\phantom{0}\nu} - 2 \bar{h}_{00} \delta^{0}_{\mu}  \delta^{0}_{\phantom{0}\nu} 
	+ 2 \delta^{0}_{\mu }  \delta^{0}_{\phantom{0}\nu} \Delta X^{a} \partial_{a} \bar{h} + 2 \Delta X^{a} \partial_{0} \bar{h}_{\mu a} \delta^{0}_{\phantom{0}\nu}  \right. \nonumber \\
	&& \ \ \ \ \ \left. + \Delta X^{a} \partial_{a} \bar{h}_{\mu \nu} + 2  \delta^{0}_{\mu} \Delta X^{a} \partial_{0} \bar{h}_{\nu a} - 2  \Delta X^{a} \partial_{a} \bar{h}_{\mu 0}  \delta^{0}_{\phantom{0}\nu} 
	-2 \delta^{a}_{\phantom{a}\nu} \Delta X_{a}   \partial_{0} \bar{h}_{\mu 0} - 2 \delta^{0}_{\mu} \Delta X^{a} \partial_{a} \bar{h}_{\nu 0} \right. \nonumber \\
	&& \ \ \ \ \   \left. -2 \delta^{a}_{\mu} \Delta X^{a}  \partial_{0}  \bar{h}_{\nu 0} + \delta^{a}_{\phantom{a}\nu} \Delta X_{a}   \partial_{\mu} \bar{h}_{00} -  \Delta X^{a} \partial_{\mu}    \bar{h}_{\nu a} 
	+ \delta^{a}_{\mu} \Delta X^{a}  \partial_{\nu}  \bar{h}_{0 0} - \Delta X^{a} \partial_{\nu}    \bar{h}_{\mu a}  \right) \label{FNC}.
\end{eqnarray}
As before, the linearized solution in eq. \eqref{VAAS_hbar} may be computed up to first order in $\GN m^2$ by iteration -- i.e., replacing all occurrences of $\bar{h}_{\mu\nu}$ on the right hand side with the first (GR) term $-16 \pi \GN \int_{x'} {}_{(1)}\overline{G}^{+} \bar{T}_{\mu\nu}$. 

{\bf Quadratic Order} \qquad Following that, we employ this order $\mathcal{O}[\GN m^2]$ accurate solution of $\bar{h}_{\mu\nu}$ in the quadratic terms of the VAAS model,
\begin{eqnarray}
	\delta_{2} \Delta G_{\mu\nu} &=&  - \frac{t^{2} m^{2}}{2} \delta_{2} G_{\mu\nu} -2 \delta_{1} V m^{2} \delta_{1} G_{\mu\nu} - \frac{1}{4} m^{2} \left( \bar{h}_{\mu\nu} - \frac{1}{2} \eta_{\mu\nu} \bar{h}   \right) \bar{h} \nonumber  \\
	&& - \frac{1}{4} m^{2} \eta_{\mu\nu}  \int_{x'} \bar{h} \ {}_{(1)}\overline{G}^{+}  \delta_{1} R + \frac{1}{2} m^{2}  \eta_{\mu\nu} \int_{x'} \delta_{1} {}_{(1)}\overline{G}^{+} \delta_{1} R + \frac{1}{2} m^{2}   \eta_{\mu\nu} \int_{x'} {}_{(1)}\overline{G}^{+} \delta_{2} R \nonumber  \\
	&& + 2 m^{2} \partial_{\mu} \partial_{\nu} \left( \delta_{2} V \right) -  2 m^{2} \eta^{\mu\lambda} \delta_{1} \Gamma^{\alpha}_{\lambda\nu} \partial_{\alpha} \left(\delta_{1} V\right) -  \eta^{\mu\lambda} \delta_{2}  \Gamma^{0}_{\lambda \nu}  (t m^{2}) \nonumber \\
	&& +t m^{2} \eta_{\mu\nu}  \partial_{0} \left(- \frac{1}{2} \int_{x'} \bar{h} \ {}_{(1)}\overline{G}^{+}  \delta_{1} R + \int_{x'} \delta_{1} {}_{(1)}\overline{G}^{+} \delta_{1} R + \int_{x'} {}_{(1)}\overline{G}^{+} \delta_{2} R   \right) \nonumber \\
	&& + \frac{1}{2} m^{2} \eta_{\mu\nu} \partial^{\rho} (\delta_{1} V) \partial_{\rho}  \bar{h} +  t m^{2} \left( \bar{h}_{\mu\nu} - \frac{1}{2} \eta_{\mu\nu} \bar{h}   \right) \partial_{0} \bar{h} \nonumber \\
	&& - \frac{t m^{2}}{2} \delta_{\{\mu}^{0} \partial_{\nu\}} \left(-\frac{1}{2} \int_{x'} \bar{h} \ {}_{(1)}\overline{G}^{+}  \delta_{1} R + \int_{x'} \delta_{1} {}_{(1)}\overline{G}^{+} \delta_{1} R + \int_{x'} {}_{(1)}\overline{G}^{+} \delta_{2} R   \right) \nonumber \\
	&& - \frac{1}{2}m^{2} \partial^{\{\mu} \delta_{1} V \partial_{\nu\}}    \bar{h} \label{eq:secondorder},
\end{eqnarray}
with the second order of Ricci tensor given by
\begin{eqnarray}
	\delta_{2} R &=& \frac{1}{8} \left(- \partial_{\alpha} \bar{h} \partial^{\alpha} \bar{h} + 8 \bar{h}^{\alpha\beta} \partial^{2} \bar{h}_{\alpha\beta} - 4 \partial_{\beta} \bar{h}_{\alpha\gamma} \partial^{\gamma} \bar{h}^{\alpha\beta} + 6 \partial_{\gamma}  \bar{h}_{\alpha\beta} \partial^{\gamma} \bar{h}^{\alpha\beta}  \right) 
\end{eqnarray}
and $\delta_1 V$ and $\delta_2 V$ already worked out in equations \eqref{VAAS_delta1V} and \eqref{VAAS_delta2V}.

Additionally, the first and second order of Christoffel symbols are
\begin{eqnarray}
	\delta_{1} \Gamma^{\alpha}_{\mu\nu} &=& \frac{1}{2} \eta^{\alpha\sigma} \left(\partial_{\mu} \left( \bar{h}_{\nu\sigma} - \frac{1}{2} \eta_{\nu\sigma} \bar{h}  \right) + \partial_{\nu} \left( \bar{h}_{\mu\sigma} - \frac{1}{2} \eta_{\mu\sigma} \bar{h} \right) - \partial_{\sigma} \left( \bar{h}_{\mu\nu} - \frac{1}{2} \eta_{\mu\nu} \bar{h} \right) \right)  \\
	\delta_{2} \Gamma^{0}_{\mu\nu} &=& \frac{1}{2} \left( \bar{h}^{0\sigma} - \frac{1}{2} \eta^{0\sigma} \bar{h} \right) \left( \partial_{\mu} \left( \bar{h}_{\nu\sigma} - \frac{1}{2} \eta_{\nu\sigma} \bar{h} \right) + \partial_{\nu} \left( \bar{h}_{\mu\sigma} - \frac{1}{2} \eta_{\mu\sigma} \bar{h} \right)  \right.  \nonumber \\
	&&  \left. \qquad \qquad \qquad \qquad - \partial_{\sigma}  \left( \bar{h}_{\mu\nu} - \frac{1}{2} \eta_{\mu\nu} \bar{h} \right) \right) . 
\end{eqnarray}
A detailed calculation leads us to discover that, the highest power of $r$ in $\delta_2 \Delta G_{\mu\nu}$ in eq. \eqref{eq:secondorder} arises from the terms
\begin{eqnarray}
	\delta_{2} \Delta G_{\mu\nu}  
	&=& 2 m^{2} (4 \pi \GN)^{2} \partial_{\mu} \partial_{\nu} \partial_{\alpha} \partial_{\beta} \Bigg\{ \int_{x'} \frac{\delta [t-t'-|\vec{x}-\vec{x}'|]}{4 \pi |\vec{x}-\vec{x}'|}  \int_{y}  \frac{\delta [t'-y^{0}-|\vec{x}'-\vec{y}|]}{4 \pi |\vec{x}'-\vec{y}|} \bar{T}^{\alpha\beta} [y] \nonumber \\
	&&\times \int_{x''} \frac{\delta [t'-t''-|\vec{x}'-\vec{x}''|]}{4 \pi |\vec{x}'-\vec{x}''|}  \int_{y'} \frac{\delta [t''-y'^{0}-|\vec{x}''-\vec{y}'|]}{4 \pi |\vec{x}''-\vec{y}'|} \left( \frac{1}{2} \bar{T}^{00} [y'] - \frac{1}{4} \bar{T} [y'] \right) \Bigg\} + \dots.  \label{leadingorderVAAS}
\end{eqnarray}
Like the DWII and ABN cases, we evaluate eq. \eqref{leadingorderVAAS} by computing its separate contributions from the near and the wave zone. We find that the near zone contribution to $(-8\pi\GN)^{-1} \delta_2 \Delta G_{\mu\nu} \subset t_{\mu\nu}$ scales as $r$ for an observer located at very large distances:
\begin{eqnarray}
	\label{VAAS_GWStressTensor_N}
	(t_{\mu\nu})_{\mathcal{N}} 
	&=& \frac{\GN m^2}{8 \pi} r \left(\delta_\mu^0 - \widehat{r}^a \delta^a_\mu \right) \left(\delta_\nu^0 - \widehat{r}^b \delta^b_\nu \right) \nonumber \\
	&\times& \partial_t^2 \Bigg\{ \int_{\vec{y}} \partial_{t} \bar{T}_{ij}[t- r,\vec{y}] \cdot
	\left( M \cdot \delta^{ij} - 2 \widehat{r}^i \widehat{r}^j 
	\left( M + \int_{\vec{y}'} \bar{T}_{ll}[t-r,\vec{y}'] \right) \right) \nonumber\\
	&&\qquad
	+ \widehat{r}^i \widehat{r}^j \int_{\vec{y}} \partial_{t}^2 \bar{T}_{ij}[t - r,\vec{y}]  
	\left( M \cdot (t - r) + \int_{-\infty}^{t - r} \int_{\vec{y}'} \bar{T}_{ll}[y'^0,\vec{y}'] \dd y'^0 \right) \nonumber\\
	&&\qquad
	+ \widehat{r}^i \widehat{r}^j \int_{\vec{y}} \bar{T}_{ij}[t - r,\vec{y}]
	\int_{\vec{y}'} \partial_{t} \bar{T}_{ll}[t - r,\vec{y}'] \Bigg\} . \label{VAAS_tmunu}
\end{eqnarray}
Turning to the wave zone contribution to eq. \eqref{leadingorderVAAS},
\begin{eqnarray}
	\label{leadingorderVAAS_WaveZone}
	(\delta_{2} \Delta G_{\mu\nu})_{\mathcal{W}}
	&\approx& \frac{1}{4} \GN^{2}  m^{2} \partial_{\mu} \partial_{\nu} \partial_{\alpha} \partial_{\beta} \int_{x'} {}_{(1)}\overline{G}^{+}  [x-x'] \frac{1}{|\vec{x}'| } \int_{\vec{y}} \bar{T}^{\alpha \beta } [t' - |\vec{x}'|, \vec{y}]  \nonumber \\
	&& \qquad \qquad \qquad \qquad 
	\left( M \cdot \left( t' - |\vec{x}'| \right) + \int^{t'  - |\vec{x}'|}_{- \infty} \text{d} y'^{0} \int_{\vec{y}'}  \bar{T}_{ll}[y'^{0}, \vec{y}'] \right).
\end{eqnarray}
Upon integrating over the $\delta-$function in ${}_{(1)}\overline{G}^{+}  [x-x']$ and comparing eq. \eqref{leadingorderVAAS_WaveZone} with equations \eqref{WaveZoneIntegral_I} and \eqref{WaveZoneIntegral_II}, we may identify
\begin{align}
	\mu^{\alpha \beta}[t', r']
	&= \frac{f^{\alpha\beta}[t'-r']}{4\pi r'} \label{WZVAAS} \\
	f^{\alpha \beta} [t'-r']
	&=  \frac{1}{2} \int_{\vec{y}} \bar{T}^{\alpha\beta}[t'-r',\vec{y}] \frac{1}{8\pi} \left( M (t'-r') + \int_{-\infty}^{t'-r'} \int_{\vec{y}'} \bar{T}_{ll}[y'^0,\vec{y}'] \dd y'^0 \right) .
\end{align}	
We therefore have $n=1$, $n'^{\langle L \rangle}=1$, and $\ell=0$, with equations \eqref{functionA} and \eqref{functionB} now producing
\begin{align}
	A[s,r] &= \int_{\mathcal{R}}^{r+s} \dd r' = r+s-\mathcal{R} \\
	B[s,r] &= \int_{s}^{r+s} \dd r' = r. 
\end{align}
Eq. \eqref{eq:wzPoisson} then returns the far zone contribution to the GW stress energy, which scales as $r^0$:
\begin{eqnarray}
	\label{VAAS_GWStressTensor_W}
	(t_{\mu\nu})_{\mathcal{W}} &=& - \frac{\GN m^2}{4 \pi} \left(\delta_\mu^0 - \widehat{r}^a \delta^a_\mu \right) \left(\delta_\nu^0 - \widehat{r}^b \delta^b_\nu \right) \nonumber \\
	&\times& \partial_t^2 \Bigg\{ \frac{M}{2} \int_{\vec{y}'} \bar{T}_{ll}[t-r,\vec{y}'] 
	+ \frac{1}{2} \widehat{r}^i \widehat{r}^j \int_{\vec{y}} \partial_t \bar{T}_{ij}[t-r,\vec{y}] \cdot \left( M(t-r) + \int_{-\infty}^{t-r} \dd y'^0 \int_{\vec{y}'} \bar{T}_{ll}[y'^0,\vec{y}'] \right) \nonumber\\
	&&\qquad
	- \widehat{r}^i \widehat{r}^j \left( \frac{M}{2} \int_{\vec{y}} \bar{T}_{ij}[t-r,\vec{y}] + \int_{\mathcal{R}}^{\infty}\dd s \int_{\vec{y}} \partial_t \bar{T}_{ij}[t-r-2s,\vec{y}] \int_{\vec{y}'} \bar{T}_{ll}[t-r-2s,\vec{y}'] \right) \nonumber\\
	&&\qquad
	+ 2 \widehat{r}^i \widehat{r}^j \left( \frac{M}{2} \int_{\vec{y}} \bar{T}_{ij}[t-r,\vec{y}] + \int_{\mathcal{R}}^{\infty}\dd s \int_{\vec{y}} \partial_t \bar{T}_{ij}[t-r-2s,\vec{y}] \int_{\vec{y}'} \bar{T}_{ll}[t-r-2s,\vec{y}'] \right)  \nonumber\\
	&&\qquad
	+ \int_{\mathcal{R}}^{\infty}\dd s  \int_{\vec{y}} \bar{T}_{ij}[t-r-2s,\vec{y}] \cdot \int_{\vec{y}'} \partial_t \bar{T}_{ll}[t-r-2s,\vec{y}'] \Bigg\} + \dots. 
\end{eqnarray}
The highest power of $r$ occurring in $\delta_2 \Delta G_{\mu\nu}$ is thus the first power $r^1$ arising from the near zone contribution in eq. \eqref{VAAS_tmunu}. In any case, we see that the GW energy flux is already ill defined.

We also checked a typical $\mathcal{O}[\GN m^2]$ nonlocal term from the quadratic piece of the Einstein tensor, which also constitutes part of the GW energy-momentum. By evaluating $\delta_2 G_{\mu\nu}$ on the linearized VAAS solution, the cross terms between the leading $m-$independent term and the first order in $m^2$ nonlocal ones will give the first GW fluxes corrected by the nonlocal interactions of the VAAS model. The term we examined was $\bar{h}^{\alpha\beta}  \partial_{\beta} \partial_{\{\mu} \bar{h}_{\nu\} \alpha}$, which in turn was dominated by the expression
\begin{eqnarray}
	\label{VAAS_GWStressTensor_Einstein}
	\bar{h}^{00}_{\text{GR}}  \partial_{0} \partial_{0} \bar{h}_{i0  (m^{2})} &\approx& - 2 m^{2}  \GN^{2} \widehat{r}^{i} M  \int_{\vec{x}''} \ddot{\bar{T}}_{ll} [t-r, \vec{x}''] + \dots \ .
\end{eqnarray}
This contribution to GW flux in eq. \eqref{GWFlux} will diverge as $r^{0+2}$, though still weaker than the $r^{1+2}$ from $(\delta_2 \Delta G_{\mu\nu})_{\mathcal{N}}$ discussed above.

\section{Summary and Discussions}
In this paper, we have investigated the gravitational wave fluxes in several nonlocal gravity models. In the DWII model, its linearized solution is the GR one multiplied by $1/(1+\bar{f})$. Minkowski spacetime is a solution to the ABN model provided certain integration constants are chosen appropriately. On the other hand, neither Minkowski $\eta_{\mu\nu}$ nor de Sitter spacetime $g_{\mu\nu}^\text{(dS)}$ solves the vacuum VAAS model because they do not render the nonlocal $\Delta G_{\mu\nu}$ proportional to the metric. It appears difficult to obtain an exact analytic solution to a vacuum cosmology in the VAAS model. Therefore, we expanded it about a vacuum cosmological spacetime, but expressed as a Fermi-Normal-Coordinate (FNC) expansion up to first order in the mass parameter squared $m^2$, where $m$ is associated with its nonlocal self-interactions.

At quadratic order, the analysis for each of the 3 nonlocal models we have examined became considerably more intricate than the one for DWI in \cite{GWFluxDWI}, though the conclusion was similar: an isolated system would generate a GW stress energy flux that appears to diverge as $r \to \infty$. More specifically, we examined the GW energy-momentum in the DWII, ABN and VAAS models by decomposing it into the near zone and the wave zone contributions \cite{gravity}; whereas for the DWI calculation in \cite{GWFluxDWI}, by focusing on the terms proportional to the second derivative of its nonlocal function $f$, only the integration over the near zone was necessary. For DWII, its near zone contribution in eq. \eqref{eq:resultnearzone} is proportional to the acceleration of the Newtonian potential, similar to the result from the first Deser-Woodard model \cite{GWFluxDWI} given by $I_{1}$, which scales as $1/r$. Apparently, we can avoid the divergent flux by setting $f'[0] \equiv \bar{f}' = 0$. Meanwhile, the nonlocal terms from both the ABN and VAAS model -- see equations \eqref{ABN_GWStressTensor_N}, \eqref{ABN_GWStressTensor_W}, \eqref{VAAS_GWStressTensor_N} and \eqref{VAAS_GWStressTensor_W} -- are unable to avoid the divergent flux $\langle \dd E/\dd t\rangle$, which scale respectively as $r^{2+2}$ and $r^{1+2}$. Furthermore, we considered the first nonlocal contributions to the quadratic order Einstein tensor $\delta_2 G_{\mu\nu}$ in the VAAS and ABN models; and found that they, too, yielded divergent gravitational fluxes. 

To be sure, the ABN and VAAS results were obtained using approximate iterative schemes in addition to the usual weak field assumption $|\bar{h}_{\mu\nu}| \ll 1$, when dealing with the nonlocal gravitational interactions. Specifically, if there were some way to solve equations \eqref{ABN_Linearized_EoM} and \eqref{VAAS_Linearized_EoM} exactly -- the latter would require knowing the full background vacuum cosmology solution -- our analysis should be re-visited. The integrals encountered while analyzing ABN and VAAS also required further regularization than those encountered in DWII; this could warrant more careful approximation schemes.

In any case, our results here reinforces that in \cite{DW I}: at the present, there are serious difficulties extracting a well defined GW energy-momentum flux from nonlocal models of modified gravity, since the usual methods do not produce the expected $1/r^2$ falloff in its momentum density $t^{0i}$. Because the quadrupole radiation formula of General Relativity is a well tested prediction, the resolution of this difficulty will hopefully lead to non-trivial physical constraints on the range of permissible nonlocal models.

\section{Acknowledgments}

YZC and AZ were supported by the Ministry of Science and Technology of the R.O.C. under the grant 106-2112-M-008-024-MY3. We thank Sohyun Park for initial collaborations on this project.

\end{document}